\documentclass[11pt]{article}

\usepackage{epsf,epsfig}
\usepackage{latexsym,rotating,amsfonts,amsmath,amssymb,float}
\usepackage{latexsym,amsfonts,amsmath,amssymb,float}
\usepackage[nolsts]{endfloat}
\usepackage{fullpage,subfigure}
\usepackage{psfrag,fancybox}
\usepackage[dvips]{color}
\usepackage{graphicx}
\usepackage[sort&compress]{natbib}

\newcommand{\pref}[1]{%
    \ref{#1} \ifnum\count0=\pageref{#1}\relax%
    \else (page \pageref{#1})\fi}

\newcommand{\eref}[1]{%
        \ref{#1}\ifnum\count0=\pageref{#1}\relax%
        \else {, p.\pageref{#1}}\fi}

\newcommand{\comment}[1]{}

\newlength{\labwidth}
\newtheorem{proposition}{Proposition}[section] 

\title{\bf Likelihood based inference for correlated diffusions}
\author{\normalfont Konstantinos Kalogeropoulos\footnote{Address:
Trumpington Street, Cambridge,UK,CB2 1PZ, Tel: +44 (0)1223 332766, E-mail: kk384@cam.ac.uk}\\
\normalfont\it University of Cambridge, Department of Engineering
- Signal Processing Laboratory
\and \normalfont Petros Dellaportas\\
\normalfont\it Athens University of Economics and Business,
Department of Statistics
\and \normalfont Gareth O. Roberts \\
\normalfont\it University of Warwick, Department of Statistics
}

\begin{document}
\maketitle

\begin{abstract}
 We address the problem of likelihood based
inference for correlated diffusion processes using Markov chain
Monte Carlo (MCMC) techniques. Such a task presents two
interesting problems. First, the construction of the MCMC scheme
should ensure that the correlation coefficients are updated
subject to the positive definite constraints of the diffusion
matrix. Second, a diffusion may only be observed at a finite set
of points and the marginal likelihood for the parameters based on
these observations is generally not available. We overcome the
first issue by using the Cholesky factorisation on the diffusion
matrix. To deal with the likelihood unavailability, we generalise
the data augmentation framework of Roberts and Stramer (2001
Biometrika 88(3):603-621) to $d-$dimensional correlated diffusions
including multivariate stochastic volatility models. Our
methodology is illustrated through simulation based experiments
and with daily EUR /USD, GBP/USD rates together with their implied
volatilities.

\bigskip


\end{abstract}

\section{Introduction}

Diffusion processes provide a natural model for phenomena evolving
continuously in time. One of their appealing features is that they
are defined in terms of the instantaneous mean and variance of the
process. Specifically, a diffusion $x_t$ obeys the dynamics of the
following stochastic differential equation (SDE)

\begin{equation}
\label{eq:standard SDE}
dx_t=\mu(t,x_t,\theta)dt+\sigma(t,x_t,\theta)dw_t,
\end{equation}

\noindent driven by standard Brownian motion $w_t$. The functions
$\mu(.)$ and $\sigma(.)$ are termed as the drift and the
volatility of the diffusion respectively. Throughout this paper we
suppress the dependence on $t$ to simplify the notation, but the
methodology is also applicable to time inhomogeneous diffusions.
The diffusion process $x_t$ is well defined if (\ref{eq:standard
SDE}) has a unique weak solution, which translates into some
regularity conditions (locally Lipschitz with a linear growth
bound) on $\mu(.)$ and $\sigma(.)$; see chapter 5 of
\cite{rog:wil94} for more details.

We address the problem of modelling several diffusions, denoted by
$x_{t}^{\{i\}},\;\;i=1,\dots,d$. Each diffusion $x_{t}^{\{i\}}$
may have a drift $\mu^{\{i\}}(.)$ and volatility
$\sigma^{\{i\}}(.)$ of general, yet known, form. We also allow for
correlations,
$corr(dx_t^{\{i\}},dx_t^{\{j\}})$=$\rho_{ij}=\rho_{ji}$, $i\neq
j$, on the instantaneous increments. The use of cross-correlations
is quite common when modelling multivariate time series, as they
may capture effects caused by common factors of the underlying
stochastic processes. In this paper we illustrate our methodology
through two examples of correlated diffusions. The first example
targets interest rates and bond pricing. Such time series often
exhibit strong inter-dependencies; for instance, interest rates
may correspond to similar bonds but with different expiry dates,
thus giving rise to correlations among them. In Section
\ref{sec:simulations} we examine a multivariate version of the
\cite{cir85} model (CIR), often used for such data. The second
example considers currency pairs which are known to be correlated,
possibly due to the common currencies they may represent. Section
\ref{sec:Real Data} contains an analysis on EUR/USD and GBP/USD
data, based on multivariate versions of stochastic volatility
diffusions, such as the model of \cite{hes93}. In both examples,
the inclusion of correlations in the model is essential for two
reasons. First, they may affect the parameter estimates of the
individual diffusions, as well as their precision. Second, they
reflect characteristics of the market which may be useful in the
bond/option pricing procedure.

We proceed by combining the diffusions $x_{t}^{\{i\}}$ together
into $X_t=(x_{t}^{\{1\}},\dots,x_{t}^{\{d\}})^{\prime}$ (with
$^{\prime}$ denoting transposition), so that $X_{t}$ is a
$d-$dimensional vector for each time t. The diffusion matrix of
$X_t$, $A$, denotes its instantaneous covariance and takes the
following form:

\begin{equation}
\label{eq:general A} A:=\left(
\begin{array}{ccccc}
\sigma^{\{1\}}(.)^2 & \rho_{12}\sigma^{\{1\}}(.)\sigma^{\{2\}}(.)
& \dots &
& \rho_{1d}\sigma^{\{1\}}(.)\sigma^{\{d\}}(.)\\
\rho_{12}\sigma^{\{1\}}(.)\sigma^{\{2\}}(.) & \sigma^{\{2\}}(.)^2
 & \dots & & \rho_{2d}\sigma^{\{2\}}(.)\sigma^{\{d\}}(.) \\
\vdots & \vdots & \ddots & & \vdots \\
\rho_{d1}\sigma^{\{1\}}(.)\sigma^{\{d\}}(.) &
\rho_{d2}\sigma^{\{2\}}(.)\sigma^{\{d\}}(.) &
 \dots & & \sigma^{\{d\}}(.)^2
\end{array}
\right)
\end{equation}

\bigskip

\noindent The diffusion process $X_t$ is defined through the
following multi-dimensional SDE
\begin{equation}
\label{eq:Multivariate SDE}
dX_t=M(X_t,\theta)dt+\Sigma(X_t,\theta)dW_t,
\end{equation}

\noindent where $W_t$ is a $d-$dimensional Brownian motion with
independent components, with vector valued drift
$M:[0,+\infty)\times \mathcal{S}_{X}\times \Theta \to \Re^{d}$
with $[M(.)]_{i}=\mu^{\{i\}}(.)$, and matrix valued volatility
(also termed as dispersion matrix)
$\Sigma(\cdot):[0,+\infty)\times \mathcal{S}_{X} \times \Theta \to
\Re^{d \times d}$, where $\mathcal{S}_{X}$ and $\Theta$ denotes
the domain of the diffusion $X_t$ and the parameter vector
$\theta$ respectively. The dispersion matrix $\Sigma$ is a
square root of the instantaneous covariance matrix
$A=\Sigma\Sigma^{\prime}$. To ensure a unique weak
solution for $X_t$, we require a unique weak solution for each
$x_t^{\{i\}}$ and the matrix $A$ to be positive definite for
all $t,X_t,\theta$.

Each diffusion $x_{t}^{\{i\}}$ may be observed, with or without
error, at a finite set of points, or may be entirely unobserved.
The diffusion will be termed as directly observed in cases with
exact observations on all $x_{t}^{\{i\}}$, and partially observed
otherwise. For ease of exposition, the methodology of this paper
is initially presented for directly observed diffusions, and
adaptations to partial observation regimes, as in multivariate
stochastic volatility models, are provided when necessary.
Similarly, we consider observations of the entire vector of $X_t$
at each time, although this assumption can easily be relaxed. We
denote the times of observations by $t_k,\;k=1,\dots,n$, and the
data with
$Y=\left\{Y_{k}=X_{t_{k}}=(x_{t_{k}}^{\{1\}},\dots,x_{t_{k}}^{\{d\}})^{\prime},
\;k=1,\dots,n\right\}$. Our aim is to draw likelihood based
inference for the parameter vector $\theta$ given these
observations.

The task of inference on diffusions observed discretely in time is
generally not trivial and has received a remarkable attention in
the recent literature; see \cite{sor04} for a recent review. The
main problem is that the likelihood is generally not available
except for a few cases. This has stimulated various techniques
based on likelihood approximations. Approximations may be
analytical \citep{ait05}, or simulation based; see \cite{ped95} or
a refinement of this technique \cite{dur:gal02}. They usually
approximate the likelihood in a way so that the discretisation
error can become arbitrarily small, although the methodology
developed in \cite{bes:pap:rob:f06} succeeds exact inference in
the sense that it allows only for Monte Carlo error.

We shall adopt a Bayesian approach using Markov
chain Monte Carlo (MCMC) method. Since diffusions are not
completely observed, it is natural to use
data augmentation \citep{tan:won87}, treating
the segments of diffusion sample path (or
a suitably fine approximation to this) as
missing data.  Initial MCMC schemes of this type
 were introduced by \cite{jon99},
\cite{era01} and \cite{ele:ch:she01}. However, as noted in the
simulation based experiment of \cite{ele:ch:she01}, and
established theoretically by \cite{rob:str01}, the algorithms
introduced in these initial implementations of MCMC in this context
degenerate as the number of imputed points increases. The
problem may be overcome for scalar diffusions with the
reparametrisation of \cite{rob:str01}. An alternative
reparametrisation is provided by \cite{gol:wil07}, see also
\cite{gol:wil06} for a sequential approach, which can in principle
be applied in principle to any diffusion.

However, the adaptation of such MCMC scheme to multivariate
diffusions introduces additional issues. The task of updating the
covariance matrix $A$ is generally not trivial, as its full
conditional posterior is most of the times intractable, and the
use of Metropolis steps is inevitable. It is therefore crucial,
especially for high-dimensional diffusions, to update the
covariance matrix componentwise as the discrepancy between
proposed and current moves is increasing in $d$. This introduces
the problem of preserving the positive definite structure of the
diffusion matrix $A$. Note that drawing samples from the
posterior of covariance matrices, which may not necessarily be
diffusion matrices, is a general MCMC issue and usually requires
appropriate matrix decompositions; see for example
\cite{pin:bat96} and \cite{dan:kas99}.

The contribution of this paper is two-fold. First, we introduce a
natural and general framework for sampling diffusion matrices in a
MCMC environment. This framework is based on the Cholesky
factorisation of $A$ and enables us to define $\Sigma $
explicitly. The MCMC algorithm may then be appropriately designed
to provide samples from the posterior of $\Sigma$, which can be
transformed to $A$ at any time through the Cholesky decomposition.
This framework may be coupled with any of the previously mentioned
likelihood approximation techniques, such as those of
\cite{bes:pap:rob:f06} or \cite{ait05}, to perform Bayesian
inference for the parameters of the multi-dimensional diffusion.
Second, we offer a full and stand alone MCMC scheme which combines
the Cholesky decomposition with the reparametrised data
augmentation approach of \cite{rob:str01}. This scheme may be used
for parameter estimation of several multivariate diffusion models
including stochastic volatility. The use of data augmentation is
justified by its convenient property to be applicable at both
directly and partially observed diffusions.

The paper is organised as follows: Section \ref{sec:reducibility}
describes the structure of a data augmentation scheme and
highlights potential problems regarding the irreducibility of the
MCMC algorithm. These problems may be tackled with the
reparametrisation of this paper which requires the Cholesky
factorisation of the diffusion matrix, presented in Section
\ref{sec:likelihood}. Specific MCMC implementation details are
given in Section \ref{sec:MCMC} and the methodology of this paper
is illustrated through simulated data in Section
\ref{sec:simulations}, and on daily EUR/USD, GBP/USD currency
pairs in Section \ref{sec:Real Data}. Finally, we summarise in
Section \ref{sec:discussion} adding some discussion and links to
some other relevant work.

\section{Data augmentation and degeneracy issues}
\label{sec:reducibility}

\subsection{The problem in practice}

Data augmentation scheme bypasses the problem of simulating directly
from the posterior $\pi(\theta|Y)$, which is typically unavailable
for discretely observed data. The idea is to introduce a
latent variable $\mathcal{X}$ that simplifies the likelihood
$\mathcal{L}(Y;\mathcal{X},\theta)$. We use the following two
steps:

\begin{enumerate}
\item \tt{Simulate $\mathcal{X}$ conditional on $Y$ and $\theta$.}

\item \tt{Simulate $\theta$ from the augmented conditional
posterior which is proportional to}\\
$\mathcal{L}(Y;\mathcal{X},\theta)\pi(\theta)$.

\end{enumerate}

Our problem can easily be adapted to this setting. $Y$ represents
the observations of the price process $X_t$, and $\mathcal{X}$
contains discrete skeletons of the diffusion paths between $Y$.
Thus, $\mathcal{X}$ and $Y$ constitute the augmented dataset
$X_{i\delta},\;i=0,\dots,T/\delta$, which is a fine partition of
the multivariate diffusion $X_t$ with $\delta$ controlling the
amount of augmentation. Based on this partition the likelihood can
be approximated, for example via the Euler-Maruyama approximation
$$
\mathcal{L}^{E}(Y;\mathcal{X},\theta)=\prod_{i=1}^{T/\delta}
p(X_{i\delta}|X_{(i-1)\delta}),
$$
\begin{equation}
\label{eq:euler} X_{i\delta}|X_{(i-1)\delta}\sim
\mathcal{N}\left(X_{(i-1)\delta}+\delta M(X_{(i-1)\delta},\theta),
\delta A(X_{(i-1)\delta},\theta)\right),
\end{equation}

\noindent which is known to converge to the true likelihood
$\mathcal{L}(Y;\mathcal{X},\theta)$ for small $\delta$
\citep{ped95}.

Another property of diffusions relates $A(X_t,\theta)$ with the
quadratic variation process. Specifically it is well-known that
\begin{equation}
\label{eq:qvar}
\lim_{\delta\rightarrow
0}\sum_{i=1}^{T/\delta}\left(X_{i\delta}-X_{(i-1)\delta}\right)
\left(X_{i\delta}-X_{(i-1)\delta}\right)^{\prime}=
\int_{0}^{T}A(X_s,\theta) ds \;\; a.s.
\end{equation}

\noindent The solution of the equation above determines the
diffusion matrix parameters exactly. Hence, there exists perfect
correlation between these parameters and $\mathcal{X}$ as
$\delta\rightarrow 0$. Thus for the theoretical algorithm which
imputes the entire $X$ path, the MCMC algorithm is reducible. In
practice this means that as the proportion of imputed data points
increases mixing problems for the MCMC chain become progressively
worse This phenomenon was first noted in \cite{rob:str01} and
\cite{ele:ch:she01}. As would be expected, the EM algorithm
suffers from the same problem.

\subsection{Measure theoretic probability viewpoint}
\label{ssec:problematic likelihood}

In this section, we explore the problem from a different angle,
through a slightly more rigorous look at the likelihood. Let $X_t$
be a diffusion that satisfies (\ref{eq:Multivariate SDE}) and
assume $X_{0}=Y_0$ and $X_{1}=Y_1$, $Y=(Y_1,Y_2)$. Denote the
probability law of $X$ by $\mathbb{P}_{\theta}$ and that of its
driftless version,

$$
d\mathcal{M}_t=\sigma(X_t,\theta)dW_t,
$$

\noindent by $\mathbb{Q}_{\theta}$. To write down the likelihood,
we can use the Cameron-Martin-Girsanov formula which provides the
Radon-Nikodym derivative of $\mathbb{P}_{\theta}$ with respect to
$\mathbb{Q}_{\theta}$:

\begin{eqnarray*}
\frac{d\mathbb{P}_{\theta}}{d\mathbb{Q}_{\theta}}\;=&G(X,M,A)&=
\;\exp\left\{ \int_0^T\left[A(X_{s},\theta)^{-1}M(X_{s},\theta)\right]^{\prime}dX_{s} \right. \\
    && - \left. \frac{1}{2}\int_{0}^{T}M(X_{s},\theta)^{\prime}A(X_{s},\theta)^{-1} M(X_{s},\theta)ds
    \right\}.
\end{eqnarray*}

\bigskip

\noindent Note that the expression above contains stochastic and
path integrals for which an analytic solution is generally not
available. However, given a sufficiently fine partition of the
diffusion path, they can be evaluated numerically providing an
approximation of the likelihood which is equivalent to
(\ref{eq:euler}).

Now assume for a moment that under $\mathbb{Q}_{\theta}$ the
marginal density of $Y$ with respect to $d-$dimensional Lebesgue
measure $Leb_{d}(Y)$, is known and denote by
$f_{\mathcal{M}}(Y;\theta)$. The dominating measure
$\mathbb{Q}_{\theta}$ can be factorised in the following way
\begin{equation}
\label{eq:measure factorisation}
\mathbb{Q}_{\theta}=\mathbb{Q}_{\theta}^{Y}\times Leb_{d}(Y)
\times f_{\mathcal{M}}(Y ;\theta),
\end{equation}

\noindent where $\mathbb{Q}_{\theta}^{Y}$ is the measure
$\mathbb{Q}_{\theta}$ conditioned on the observations $Y$. We can
now write
\begin{equation}
\label{eq:likelihood1}
\frac{d\mathbb{P}_{\theta}}{\mathbb{Q}_{\theta}^{Y}\times
Leb_{d}(Y)}(X^{mis},Y)\;=\;G(X,M,A)\times f_{\mathcal{M}}(Y
;\theta).
\end{equation}

The expression in (\ref{eq:likelihood1}) provides the likelihood
for the latent diffusion paths $X^{mis}$ and the parameters
$\theta$. However, this likelihood is not valid because its
reference measure, $\mathbb{Q}_{\theta}^{y}$, depends on
parameters. Furthermore, since the volatility parameters are
identified by the quadratic covariation process, the measure
$\mathbb{Q}_{\theta}$ is just a point mass. Consequently, the
measures $\mathbb{Q}_{\theta}$ are mutually singular and therefore
so are $\mathbb{P}_{\theta}$. Hence, inference for both
$X^{mis},\theta$ is not possible using a common $\sigma-$finite
dominating measure. In the next section, we specify an appropriate
transformation of the diffusion that allows a likelihood
specification with respect to a parameter-free dominating measure.
This transformation may be viewed as a generalisation of the one
in \cite{rob:str01}. The transformed diffusion has unit
volatility, thus the problems induced by the quadratic variation
property of (\ref{eq:qvar}) are implicitly addressed.

\section{Likelihood specification}
\label{sec:likelihood}

\subsection{A Cholesky factorisation of the diffusion matrix}
\label{ssec:cholesky}

Consider the multi-dimensional SDE of (\ref{eq:Multivariate SDE})
with the diffusion matrix $A$ of (\ref{eq:general A}). The
$d\times d$ matrices $A$ and $\Sigma$ are linked through
$A=\Sigma\Sigma^{\prime}$, therefore $\Sigma$ is not unique.
However, it is crucial to define $\Sigma$ explicitly and
establish a 1-1 mapping with $A$, as each one of these two
matrices may be more convenient for different reasons. The
likelihood, defined either through the Euler-Maruyama
approximation in (\ref{eq:euler}) or through
Cameron-Martin-Girsanov's formula in (\ref{eq:likelihood1}), is
expressed in terms of $A$, which is also the main target of
inference. On the other hand $A$ is a positive definite matrix,
whereas the only assumption made on $\Sigma$ requires its full
rank. Hence it is generally more convenient to work with
$\Sigma$ in the context of a MCMC algorithm. Moreover, as
mentioned in the previous section, the generalisation of the
\cite{rob:str01} reparametrisation involves a transformation to
unit volatility which will naturally be based on $\Sigma$.

In this paper, we define $\Sigma$ using the Cholesky
decomposition of $A$. Let
$S_{x}(X_t,\theta)=diag\{\sigma^{\{i\}}(X_t,\theta)\}$. The
diffusion matrix may then be factorised in the following way
$$
A(X_t,\theta)\;=\;S_{x}(X_t,\theta)\;R\;S_{x}(X_t,\theta),
$$

\noindent where $R$ is the correlation matrix. One may define
$\Sigma$ as the product of $S_x$ with the Cholesky decomposition
of R, say C. But the elements of C will not have the general
Cholesky structure, since R has the additional property of being a
correlation matrix. To eliminate such problems we write each
$\sigma_{i}(X_t,\theta)$ as
\begin{equation}
\label{eq:rescale}
\sigma^{\{i\}}(X_t,\theta)=c_{i}f^{\{i\}}(X_t,\theta),\;\;\forall
i,
\end{equation}

\noindent for some positive constants $c_{i}$. This imposes no
restrictions as we can always set
$f^{\{i\}}(X_t,\theta)=\sigma^{\{i\}}(X_t,\theta)/c_{i}$, see
Section \ref{ssec:multivariate sv} for such an example. Now, based
on $F_{x}(X_t,\theta)=diag\{f^{\{i\}}(X_t,\theta)\}$, we can use
(\ref{eq:rescale}) to obtain an alternative decomposition of
$A$,
$$
A(X_t,\theta)\;=\;F_{x}(X_t,\theta)\;V\;F_{x}(X_t,\theta),
$$

\noindent where $V$ is a general symmetric positive definite
matrix with
\begin{equation}
\label{eq:V matrix}
V_{ij}=\left\{\begin{array}{cc}
c_{i}^2,\;&  i=j\\
\rho_{ij}c_{i}c_{j},& i\neq j. \end{array}\right.
\end{equation}

\noindent The Cholesky decomposition of $V$, denoted by $C$
($V=CC^{\prime}$), may now be used. The dispersion matrix
$\Sigma(X_t,\theta)$ is defined as
\begin{equation}
\label{eq:cholesky matrix}
\Sigma(X_t,\theta)\;=\;F_{x}(X_t,\theta)\;C.
\end{equation}

\noindent In coordinate form, $\Sigma$ may be written as
$$
[\Sigma(X_t,\theta)]_{ij}=\left\{\begin{array}{cc}
[C]_{ij}f_{i}(X_t,\theta),&  j\leq i\\
0,& j>i. \end{array}\right.
$$

\noindent The only restriction on the constants $C_{ij}$ requires
compatibility with the Cholesky decomposition, which translates on
positive diagonal entries $C_{ii}$. As we mention in
\ref{ssec:updates sigma}, this is particularly convenient in a
MCMC environment and specifically for componentwise updates of
$\Sigma(X_t,\theta)$ parameters. The Cholesky decomposition
establishes the 1-1 mapping between $\Sigma$ and $A$ and
ensures that the entire space of diffusion matrices as $A$ is
covered.

\subsection{Transformation to unit volatility}
\label{ssec:transformation}

In Section \ref{sec:reducibility}, the need for a
reparametrisation was highlighted in order to avoid degenerate
MCMC algorithms. \cite{rob:str01} provide a solution to the
problem for scalar diffusions, which involves a transformation to
unit volatility. However, in more than one dimensions such a
transformation does not always exist, as noted \cite{ait05}. When
such a transformation is available the diffusion is said to be
reducible, a term introduced by \cite{ait05} who also provides a
necessary and sufficient condition for reducibility: diffusions
with non-singular $\Sigma(X_t,\theta)$ are reducible if and only
if
\begin{equation}
\label{eq:sahalia condition}
\frac{\partial
[\Sigma(X_t,\theta)^{-1}]_{ij}}{\partial x_{t}^{\{k\}}}=
\frac{\partial [\Sigma(X_t,\theta)^{-1}]_{ik}}{\partial
x_{t}^{\{j\}}},\;\forall\; i,j,k\in\{1,\dots,d\},\;\text{with }j<k
\end{equation}

Not all SDEs with diffusion matrix $A$ as in (\ref{eq:general A})
or dispersion matrix $\Sigma$ as in (\ref{eq:cholesky matrix}) are
reducible. In this section, we restrict our attention to
diffusions with
\begin{equation}
\label{eq:reducible class} \sigma^{\{i\}}(X_t,\theta)\equiv
\sigma^{\{i\}}(x_t^{\{i\}},\theta),
\end{equation}

\noindent for which we prove the reducibility. This is established
by the following proposition:

\proposition \label{prop:reducibility} Let X be a $d$-dimensional
diffusion which obeys the following SDE:
$$
dX_t=M(t,X_t,\theta)dt+\Sigma(t,X_t,\theta)dW_t.
$$

\noindent Furthermore, assume that
$$
\Sigma(X_t,\theta)\;=\;F_{x}(X_t,\theta)\;C,
$$
\noindent where
$F_{x}(X_t,\theta)=diag\{f^{\{i\}}(x_t^{\{i\}},\theta)\}$ and C is
a lower triangular matrix with positive diagonal elements. The
diffusion X can then be transformed to one with identity diffusion
matrix. In other words X is reducible.

\bigskip

\noindent \normalfont {\bf Proof:} See Appendix.

\bigskip

\noindent The next proposition provides explicitly a
transformation to unit volatility. It may be viewed as an
alternative proof of proposition \ref{prop:reducibility}

\proposition \label{prop:unit vol transformation} Consider the
setting and the diffusion $X_t$ of proposition
\ref{prop:reducibility}. Suppose that there exist
$g^{\{i\}}(x_{t}{\{i\}},\theta)$ for $i=1,\dots,d$ with continuous
second derivatives, so that
$$
\frac{\partial g^{\{i\}}(x_{t}^{\{i\}},\theta)}{\partial
x_{t}^{\{i\}}}= \frac{1}{f^{\{i\}}(x_{t}^{\{i\}},\theta)},\;
j=1,\dots,d,
$$

\noindent and let
$G_x(X_t,\theta)=\left(g^{\{1\}}x_{t}^{\{1\}},\theta),
\dots,g^{\{d\}}(x_{t}^{\{d\}},\theta)\right)^{\prime}$. Consider
the transformation
\begin{equation}
\label{eq:unit vol transformation}
H(X_t,\theta)=\left(h^{\{1\}}(X_t,\theta),\dots,h^{\{d\}}(X_t,\theta)\right)^{\prime}=C^{-1}G_x(X_t,\theta).
\end{equation}

\noindent The diffusion $U_t=H(X_t,\theta)$ has then unit
volatility.

\bigskip

\noindent \normalfont {\bf Proof:} See Appendix.

\bigskip  The transformation of (\ref{eq:unit vol transformation}) may be used to
specify the likelihood under an appropriate reparametrisation
which will ensure a non - decreasing efficiency, of the data
augmentation MCMC scheme, in the level of augmentation. Notice
that the transformation of (\ref{eq:unit vol transformation}) to
unit volatility is not unique. This is not necessary for our
methodology, in fact we only require its invertibility which is
ensured as long as each $g_{i}(x_{t}{\{i\}},\theta)$ is itself
invertible. We present this reparametrisation in the Section
\ref{ssec:likelihood}, whereas in \ref{ssec:multivariate sv} we
show how to relax the assumption of (\ref{eq:reducible class}) to
handle multivariate stochastic volatility models.

\subsection{Reparametrised likelihood}
\label{ssec:likelihood}

Consider the diffusion that satisfies the SDE of
(\ref{eq:Multivariate SDE}) where the drift $M(.)$ and $\Sigma$
satisfy the appropriate conditions so that $X_t$ has a unique weak
solution and Ito's lemma can be applied. Furthermore, assume that
$$
\Sigma(X_t,\theta)\;=\;F_{x}(X_t,\theta)\;C,
$$
\noindent where
$F_{x}(X_t,\theta)=diag\{f^{\{i\}}(x_t^{\{i\}},\theta)\}$ and C is
a lower triangular matrix with positive diagonal elements. For
ease of illustration let the entire vector of $X_t$ be observed at
each time and denote the times of observations by
$t_k,\;k=0,\dots,n$, and the data with
$Y=\left\{Y_{k}=X_{t_{k}}=(x_{t_{k}}^{\{1\}},\dots,x_{t_{k}}^{\{d\}})^{\prime},
\;k=1,\dots,n\right\}$. We will define the likelihood for a pair
of successive observations, ($Y_{k-1},Y_{k}$). Due to the Markov
property of diffusions, the full likelihood is just given by the
product of all pairs of consecutive observations. Without applying
a reparametrisation, the likelihood can be defined through
(\ref{eq:likelihood1}). However, as discussed in
\ref{sec:reducibility}, this likelihood is problematic because it
is written with respect to a dominating measure that depends on
parameters. The aim of the reparametrisation is to obtain a
likelihood with a parameter-free dominating measure.

The first step of the reparametrisation requires a transformation
$U_t=H(X_t,\theta)$ =
$\left(u^{\{1\}},\dots,u^{\{d\}}\right)^{\prime}$, so that the
diffusion matrix of $U_t$ is the $d-$dimensional identity matrix.
As established by proposition \ref{prop:reducibility}, such a
transformation does exist and can be obtained explicitly by
(\ref{eq:unit vol transformation}). The SDE of the $r-$th
coordinate of the transformed diffusion $U$ will be given by:
$$
du_{t}^{\{r\}}=\mu_{U}^{\{r\}}(U_t,\theta)dt+ dw_{t}^{\{r\}},\;
r=1,\dots,d,
$$

\noindent with
$$
\mu_{U}^{\{r\}}(U_t,\theta)=\sum_{i=1}^{d}\frac{\partial
h_{r}(X_t,\theta)}{\partial
x^{\{i\}}}\mu^{\{i\}}(X_t,\theta)+\sum_{i=1}^{d}\frac{\partial^2
h_{r}(X_t,\theta)}{\partial
(x^{\{i\}})^2}[\Sigma(X_t,\theta)]_{ii}^2,
$$

\noindent where $X_t$ may replaced with $H^{-1}(U_t,\theta)$ so
that the SDE is expressed in terms of $U_t$. If we use the
Cameron-Martin-Girsanov formula in a similar manner as in Section
\ref{ssec:problematic likelihood}, we can write the likelihood as
$$
\frac{d\mathbb{P}_{\theta}}{\mathbb{W}^{Y^{H}}\times
Leb_{d}(Y^{H})}\left(U^{mis},Y\right)\;=\;G(U,\mu_{U},I_d)
f_{\mathcal{M}}(Y ;\theta),
$$

\noindent or equivalently
$$
\frac{d\mathbb{P}_{\theta}}{\mathbb{W}^{Y^{H}}\times
Leb_{d}(Y)}\left(U^{mis},Y\right)\;=\;G(U,\mu_{U},I_d)\times
\mathcal{N}\left(Y^{H}_k-Y^{H}_{k-1},I_{d}\right)|J(Y,\theta)|,
$$

\noindent where $\mathbb{W}^{Y^{H}}$ is just Wiener measure
conditioned on the transformed observations $Y^{H}$=$H(Y,\theta)$,
$\mathcal{N}(Y,V)$ denotes the Gaussian density of $Y$ under 0
mean and covariance V, and $J(Y,\theta)$ is the Jacobian term from
the transformation $H(Y,\theta)$. The dominating measure of the
likelihood, $\mathbb{W}^{Y^{H}}$, reflects the distribution of $d$
independent Brownian bridges with $Y^{H}$ as endpoints and
therefore depends on parameters. For this reason we introduce a
second transformation
\begin{equation}
\label{eq:2nd repar} z^{\{i\}}(s)=u^{\{i\}}(s)-
\frac{(t_{k}-s)H(y_{k-1}^{\{i\}},\theta)(t_{k-1})+(s-t_{k-1})h(y_{k}^{\{i\}},\theta)}{t_{k}-t_{k-1}},
\: t_{k-1}<s<t_{k},
\end{equation}

\noindent for all $i\in\{1,\dots,d\}$, which centers the bridge to
start and finish at 0 and preserves the unit volatility. Let
$Z=\left(z^{\{1\}},\dots,z^{\{d\}}\right)^{\prime}$and the
function $U=\eta(Z)$ to be the inverse of \ref{eq:2nd repar}. The
SDE for $Z$ becomes
$$
dz_t^{\{i\}}=\mu_{U_t}^{\{i\}}(\eta(Z_t),\theta)dt+dw_t^{\{i\}},\;\forall\;i\in\{1,\dots,d\}
$$

\noindent The likelihood may now be written as
\begin{equation}
\label{eq:repar likelihood}
\frac{d\mathbb{P}_{\theta}}{\mathbb{W}^{0)}\times
Leb_{d}(Y)}\left(Z^{mis},h(Y,\theta)\right)\;=\;G(\eta(Z_t),M_{U},I_d)\times
\mathcal{N}\left(Y^{H}_{k}-Y^{H}_{k-1},I_{d}\right)|J(Y,\theta)|,
\end{equation}

\noindent where
$$
M_U=\left(\mu_{U_t}^{\{1\}}(\eta(Z_t),\theta),\dots,\mu_{U_t}^{\{d\}}(\eta(Z_t),\theta)\right)^{\prime}.
$$

\noindent The dominating measure of the likelihood provided by
\ref{eq:repar likelihood} does not depend on any parameters, being
the product of $d$ independent Brownian bridges that start and
finish at 0. The likelihood of (\ref{eq:repar likelihood}) may be
used to construct an irreducible MCMC scheme which will not
degenerate as we increase the amount of augmentation. The
stochastic and path integrals involved cannot be solved
analytically but they can be evaluated numerically given a
sufficiently fine partition of the diffusion path. Note also that,
as a result of these transformations, inference will now be based
on $Z_t$ rather than $X_t$. However, the posterior draws of $Z_t$
may be inverted to provide samples from the posterior of $X_t$.

\subsection{Multivariate stochastic volatility models}
\label{ssec:multivariate sv}

In the previous subsection we assumed a diffusion with SDE that
satisfies (\ref{eq:reducible class}) so that the transformation of
(\ref{eq:unit vol transformation}) is directly applicable.
However, there exist interesting diffusion models outside of this
class with a broad range of applications. One famous example of
such models is provided by stochastic volatility; see for example
\cite{ghy:har:ren96}. Most diffusion driven stochastic volatility
models, including those of \cite{hul:whi87}, \cite{ste:ste91} and
\cite{hes93}, belong to the following general class of
$2-$dimensional SDEs
\begin{equation}
\label{eq:sv models}
\left(\begin{array}{ccc}dx_{t}\\dv_{t}\end{array}\right)=
\left(\begin{array}{ccc}\mu_x(v_{t},\theta)\\
\mu_v(v_{t},\theta)\end{array}\right)dt
+\left(\begin{array}{ccc}\sigma_x(v_{t},\theta) & 0\\
0 & \sigma_v(v_{t},\theta)\end{array}\right)\left(\begin{array}{ccc}db_{t}\\
dw_{t}\end{array}\right),
\end{equation}

\noindent where $b_t$ and $w_t$ are correlated standard Brownian
motions, $x_{t}$ usually denotes the log price, whose volatility
is provided by another diffusion $v_{t}$.

Diffusions that satisfy SDEs as in (\ref{eq:sv models}) cannot
generally be transformed to unit volatility \citep{ait05}, as the
reparametrisation of \ref{ssec:likelihood} requires. Nevertheless,
it is still possible to construct an irreducible data augmentation
scheme to estimate their parameters. As noted in
\cite{ch:pit:she05} the conditional likelihood of $x_{t}$, given
$v_{t}$, is available in closed form and therefore only the paths
of $v_{t}$ need to be imputed to approximate the likelihood.
Consequently, as shown in \cite{kal07}, it suffices to transform
$v_{t}$ itself to unit volatility.

This idea may be coupled with the Cholesky factorisation to handle
multivariate stochastic volatility models. We illustrate this for
the case of a bivariate Heston model. The scalar Heston model can
be written as
\begin{eqnarray*}
dx_{t}&=&\left(\mu_x-\frac{1}{2}v_{t}^2\right)dt+\sqrt{v_{t}}db_t,\\
dv_{t}&=&\kappa\left(\mu_v-v_{t}\right)dt+\sigma\sqrt{v_{t}}dw_t.
\end{eqnarray*}

\noindent where $b_t$ and $w_t$ are correlated. We can re-write
the top equation, by setting $c=\sqrt{\mu_v}$, to
$$
dx_{t}=\left(\mu_x-\frac{1}{2}v_{t}^2\right)dt+c\sqrt{\frac{v_{t}}{\mu_v}}dB_t.
$$

\noindent Based on the formulation above, a bivariate Heston model
may be written as a $4-$dimensional diffusion
$X_t=\left(v_{t}^{\{1\}},v_{t}^{\{2\}},x_{t}^{\{1\}},x_{t}^{\{2\}}\right)^{\prime}$,
with $x_{t}^{\{1\}},x_{t}^{\{2\}}$ denoting the log-prices, and
$v_{t}^{\{1\}},v_{t}^{\{2\}}$ their volatilities. The diffusion
matrix now has the general form of (\ref{eq:general A}) all of the
components of $X_t$ may be correlated. Since (\ref{eq:rescale})
holds for each component of $X_t$, we can define the dispersion
matrix of $X_t$ as in (\ref{eq:cholesky matrix})

\begin{equation}
\label{eq:bivariate heston}
\left(\begin{array}{ccc}dv_{t}^{\{1\}}\\dv_{t}^{\{2\}}
\\dx_{t}^{\{1\}}\\dx_{t}^{\{2\}}\end{array}\right)=
\left(\begin{array}{ccc}\kappa_1\left(\mu_1-v_{t}^{\{1\}}\right)\\
\kappa_2\left(\mu_2-v_{t}^{\{2\}}\right)\\
\mu_3-\frac{1}{2}(v_{t}^{\{1\}})^2\\
\mu_4-\frac{1}{2}(v_{t}^{\{2\}})^2\end{array}\right)dt\;+
\;F_x(X_t,\theta)\;C\;dB_t,
\end{equation}

\noindent where now $B_t$ is a $4-$dimensional Brownian motion
with independent components,
$$
F_x(X_t,\theta)=diag\left\{\sqrt{v_{t}^{\{1\}}},\sqrt{v_{t}^{\{2\}}},
\frac{\sqrt{v_{t}^{\{1\}}}}{\mu_1},\frac{\sqrt{v_{t}^{\{2\}}}}{\mu_2}\right\},
$$

\noindent and $C$ is the lower triangular Cholesky matrix whose
entries $C_{ij}$ may be seen as a 1-1 transformation of parameter
vector containing the correlations $\rho_{ij}$, and also
$\sigma_1$, $\sigma_2$, $\sqrt{\mu_1}$ and $\sqrt{\mu_2}$.

Regarding the likelihood, consider again a pair of successive
observations, $Y_{k-1},Y_{k}$ with
$Y_{k}=(y_{k}^{\{3\}},y_{k}^{\{4\}})$, for
$x_{t}^{\{1\}},x_{t}^{\{2\}}$. Conditional on
$v_{t}^{\{1\}},v_{t}^{\{2\}}$, and therefore also on their
corresponding Brownian components $b_{t}^{\{1\}},b_{t}^{\{2\}}$,
the likelihood for $Y_{k}$ is a bi-variate Gaussian with mean
$$
\left(\begin{array}{ccc}
y_{k-1}^{\{3\}}+\int_{t_{k-1}}^{t_k}\left(\mu_3-\frac{1}{2}(v_{s}^{\{1\}})^2\right)ds+
C_{31}\int_{t_{k-1}}^{t_k}\sqrt{\frac{v_{s}^{\{1\}}}{\mu_1}}db_{s}^{\{1\}}+
C_{32}\int_{t_{k-1}}^{t_k}\sqrt{\frac{v_{s}^{\{1\}}}{\mu_1}}db_{s}^{\{2\}}\\
y_{k-1}^{\{4\}}+\int_{t_{k-1}}^{t_k}\left(\mu_4-\frac{1}{2}(v_{s}^{\{2\}})^2\right)ds+
C_{41}\int_{t_{k-1}}^{t_k}\sqrt{\frac{v_{s}^{\{2\}}}{\mu_2}}db_{s}^{\{1\}}+
C_{42}\int_{t_{k-1}}^{t_k}\sqrt{\frac{v_{s}^{\{2\}}}{\mu_2}}db_{s}^{\{2\}}
\end{array}\right),
$$

\noindent and covariance matrix
$$
\left(\begin{array}{ccc}
\int_{t_{k-1}}^{t_k}C_{33}^2\frac{v_{s}^{\{1\}}}{\mu_{3}^2}ds &
\int_{t_{k-1}}^{t_k}C_{33}C_{43}\frac{\sqrt{v_{s}^{\{1\}}v_{s}^{\{2\}}}}{\mu_{3}\mu_{4}}ds \\
\int_{t_{k-1}}^{t_k}C_{33}C_{43}\frac{\sqrt{v_{s}^{\{1\}}v_{s}^{\{2\}}}}{\mu_{3}\mu_{4}}ds&
\int_{t_{k-1}}^{t_k}(C_{43}^2+C_{44}^2)\frac{v_{s}^{\{2\}}}{\mu_{4}^2}ds
\end{array}\right).
$$

\noindent The integrals above cannot be computed analytically, but
the augmented path of $v_{t}^{\{1\}},v_{t}^{\{2\}}$ enables
accurate numerical approximations of them.

The remaining part of the likelihood may be obtained through the
reparametrisation recipe of Section \ref{ssec:likelihood},
modified according to the observation regime of the volatility. In
some cases the volatility may be entirely unobserved, leading to a
partially observed diffusion. Nevertheless alternative
formulations are available, where information from option prices
is used to construct exact or noisy volatility observations; see
for example \cite{ait:kim05}, \cite{che:ghy00} and
\cite{kal:rob:del07}. In the presence of exact observations the
transformations of (\ref{eq:unit vol transformation}) and
(\ref{eq:2nd repar}) may be used. Note that transformation to unit
volatility refers to the 2-dimensional diffusion
$(v_{t}^{\{1\}},v_{t}^{\{2\}})^{\prime}$, rather than the entire
$X_t$. For the bivariate Heston model it takes the following form
$$
U_t=H(X_t,D)=D^{-1}G_x(X_t),
$$
\noindent where
$$
G_x(X_t)=\left(2\sqrt{x_t^{\{1\}}},
2\sqrt{x_t^{\{2\}}}\right)^{\prime},
$$
\noindent and $D$ is a block of $C$ containing the $C_{ij}$
entries with $i,j=\{1,2\}$. If the observations are noisy or they
do not exist at all, the transformation of (\ref{eq:2nd repar})
may be replaced with
$$
Z^{\{i\}}(s)=U^{\{i\}}(s)-U_0, \: 0<s<t_{n},
$$

\noindent and the
$\mathcal{N}\left(Y^{H}_k-Y^{H}_{k-1},I_{d}\right)|J(Y,\theta)|$
part of the likelihood should be replaced with the relative noise
density or removed accordingly.

The above likelihood specification can be applied to all
multivariate stochastic volatility models that satisfy the SDE of
\ref{eq:sv models}. For more complex models, the framework of
\cite{gol:wil07} or time change transformations of
\cite{kal:rob:del07} may be combined with the Cholesky
factorisation.

\section{MCMC implementation}
\label{sec:MCMC}

Based on the likelihood specifications of the previous section, it
is now possible to construct an irreducible data augmentation MCMC
scheme. The algorithm may be divided into three parts: the updates
of the diffusion paths $Z^{mis}$, the parameters of the dispersion
matrix $\Sigma(X_t,\theta)$ and those of the drift
$M(X_t,\theta)$. Generally, the updates of the drift parameters
may be executed using standard random walk Metropolis techniques,
although for some diffusion models the full conditionals may be
analytically tractable and Gibbs steps may be used instead. Hence,
in the next two subsections we provide some details regarding the
updates of the diffusion paths and the volatility parameters.

\subsection{Updating the imputed paths}

There exist several options for carrying out this step and most of
them are based on an independence sampler. For discretely observed
diffusions the augmented path may be divided into $n\times d$
diffusion bridges connecting the observed points, and each one of
them may be updated in turn. The full conditional of $Z^{mis}$ may
be written as
\begin{equation}
\label{eq:ind sampler1}
\frac{d\mathbb{P}_{\theta}}{d\mathbb{W}^{0}}(Z^{mis}|Y)=
G(\eta(Z_t),M_{U},I_d)\frac{f_{\mathcal{M}}(Y;A)}{f_{\mathcal{X}}(Y;A)}
\propto G(\eta(Z_t),M_{U},I_d),
\end{equation}

\noindent where $f_{\mathcal{X}}(Y;A)$ is the density of $Y$ with
respect to the Lebesgue measure under $\mathbb{P}_{\theta}$. Note
that this expression will be slightly different for stochastic
volatility models.

The dominating measure of the likelihood $\mathbb{W}^{0}$, in
other words a Brownian bridge, may be used as the proposal
distribution for the independence sampler. Based on (\ref{eq:ind
sampler1}), the algorithm will then contain the following steps

\begin{itemize}
\tt{

\item{Step 1:} Propose a Brownian bridge from $t_{k-1}$ to
$t_{k}$.

\item{Step 2:} Substitute into $i$-th dimension and form $Z_t^*$.

\item{Step 3:} Accept with probability:
$$
\min\left\{1,\frac{G(\eta(Z_t^*),M_{U},I_d)}{G(\eta(Z_t),M_{U},I_d)}\right\}.
$$

\item Repeat for all $k=1,\dots n$ and $i=1,\dots,d$. }

\end{itemize}

\noindent The algorithm above takes advantage of the
transformation to unit volatility and splits the path into
$n\times d$ independent, under the dominating measure, bridges.
Alternative proposals are available such as the diffusion bridges
introduced in \cite{dur:gal02} and \cite{del:hu07}, which can be
adapted in a MCMC setting through the reparametrisation framework
of \cite{gol:wil07}. Another option is to propose local moves of
the paths in the spirit of \cite{bes:rob:stu:vos06}. This approach
may be viewed as a random walk metropolis in the space of
diffusion bridges. Note however that this technique requires
bridges with unit volatility, and therefore it can only be used
for correlated diffusions through the reparametrisation framework
of this paper.

Further increase in the acceptance rate may be achieved by
choosing a proposal distribution which is closer to the target
$\mathbb{P}_{\theta}$, for example a linear diffusion bridge.
Suppose that we propose from another diffusion bridge
distribution, denoted by $\mathbb{L}^{0}$, with drift $L$. We can
now write:
\begin{equation}
\label{eq:ind sampler2}
\frac{d\mathbb{P}_{\theta}}{d\mathbb{L}^{0}}(Z_{mis}|Y)=
\frac{d\mathbb{P}_{\theta}/d\mathbb{W}^{0}}
{d\mathbb{L}_{0}/d\mathbb{W}^{0}}(Z_{mis}|Y) \propto
\frac{G(\eta(Z_t),M_{U},I_d)}{G(\eta(Z_t),L,I_d)}
\end{equation}

\noindent Based on (\ref{eq:ind sampler2}), the corresponding
algorithm, termed as method B in \cite{rob:str01}, will consist of
the following steps:

\begin{itemize}
\tt{

\item{Step 1:} Propose a Brownian bridge from $t_{k-1}$ to
$t_{k}$.

\item{Step 2:} Substitute into $i$-th dimension and form $Z_t^*$.

\item{Step 3:} Accept with probability:
$$
\min\left\{1,\frac{G(\eta(Z_t^*),M_{U},I_d)G(\eta(Z_t),L,I_d)}
{G(\eta(Z_t^*),L,I_d)G(\eta(Z_t),M_{U},I_d)}\right\}.
$$

\item Repeat for all $k=1,\dots n$ and $i=1,\dots,d$. }

\end{itemize}

However, low acceptance rates may still occur, especially in
sparse datasets. In such cases, each bridge may be further split
into smaller blocks and updating strategies based on overlapping
or random sized blocks may be advocated; see \cite{kal07} and
\cite{ch:pit:she05} for more details. These techniques may also be
used in partially observed diffusions, for example in stochastic
volatility models, where some components of the diffusion may be
observed with error or not be observed at all.

\subsection{Updating the volatility parameters}
\label{ssec:updates sigma}

As mentioned earlier, the parameter updates of the diffusion
matrix $A(X_t,\theta)$ are not trivial. Their full conditional
posterior is generally not available in closed form, and
Metropolis steps are inevitable. The construction of such steps
has to ensure that the covariance matrix structure of
$A(X_t,\theta)$ is preserved. At the same time, it is desirable to
achieve a reasonably high acceptance rate of the proposed moves
for a good mixing of the MCMC algorithm. While the former may be
implemented by using an appropriate distribution for symmetric
positive definite matrices, such as the Wishart distribution, it
is extremely difficult to guarantee the latter, especially for
high dimensional diffusions.

The Cholesky factorisation introduced in this paper may be of help
in such cases. Specifically, the step of updating the constants
$c_{i}$, and the correlations $\rho_{ij}$, with $i,j
\in\{1,\dots,d\}$ and $i<j$, may be replaced by componentwise
updates of the Cholesky matrix $C$. In contrast with the
correlations $\rho_{ij}$, the restrictions implied by the
symmetric and positive definite diffusion matrix $A(X_t,\theta)$
may be enforced on the elements of C in a straightforward manner,
as only the positivity of the diagonal entries is required.

Hence, the updates of $C_{ij}$'s may be implemented through
standard random walk Metropolis steps. Note that
($c_{i}$,$\rho_{ij}$) and $C_{ij}$ are linked through
\begin{equation}
\label{eq:system}
S_{x}(X_t,\theta)\;R\;S_{x}(X_t,\theta)\;=
\;F_{x}(X_t,\theta)\;V\;F_{x}(X_t,\theta)\;=\;A(X_t,\theta),
\end{equation}

\noindent where $R$ is the correlation matrix and $V$ is defined
in (\ref{eq:V matrix}). It is not hard to see that they are linked
with an 1-1 mapping which is the solution of the system in
(\ref{eq:system}) with $d(d+1)/2$ equations and unknowns. Hence,
the draws from the posterior of $C$ may be transformed back at any
time, to obtain draws from the posterior of ($c_{i}$,$\rho_{ij}$).

\section{Simulation based experiments}
\label{sec:simulations}

In this section we illustrate and test our data augmentation
scheme on a $3-$dimensional CIR model. In other words, we consider
a $3-$dimensional diffusion
$X_t=(x_{t}^{\{1\}},x_{t}^{\{2\}},x_{t}^{\{3\}})^{\prime}$ with
linear drift for each component $\kappa_{i}(\mu_{i}-x_t^{\{i\}})$,
the CIR formulation of the volatility,
$\sigma_i\sqrt{x_t^{\{i\}}}$, and correlations between all the
components, $\rho_{ij}$, $i=1,2,3$, $j<i$. This model may be
useful for the analysis of interest rates time series, where the
cross-correlations may be substantial. Notice that our framework
allows for more general drift and volatility formulations but the
main focus of this simulation experiment lies mainly in the
correlations $\rho_{ij}$. The dispersion matrix of the
multi-dimensional diffusion $X_t$ may be defined as in
(\ref{eq:cholesky matrix}), with
$$
F_x(X_t,\theta)=diag\left\{\sqrt{x_t^{\{1\}}},\sqrt{x_t^{\{2\}}},\sqrt{x_t^{\{3\}}}\right\},
$$

\noindent and $C$ being the lower triangular matrix from the
Cholesky decomposition, whose entries $C_{ij}$, substitute the
parameters $\sigma_i$ and $\rho_{ij}$. The likelihood
reparametrisation requires a transformation to unit volatility
which is given by
$$
U_t=H(X_t,C)=C^{-1}G_x(X_t),
$$
\noindent with
$$
G_x(X_t)=\left(2\sqrt{x_t^{\{1\}}}, 2\sqrt{x_t^{\{2\}}},
2\sqrt{x_t^{\{3\}}}\right)^{\prime}.
$$

\noindent The second transformation is that of (\ref{eq:2nd
repar}), and the likelihood may be obtained from (\ref{eq:repar
likelihood}). To complete the model formulation we assign
non-informative priors: $p(\theta)\propto \theta^{-1}$ for the
positive parameters $\kappa_{i},\mu_{i},C_{ii}$ and
$p(\theta)\propto 1$ for the rest ($C_{ij},i>j$).

\begin{figure}

\psfrag{Lag - C11}{Lag - $C_{11}$}

\psfrag{Lag - C21}{Lag - $C_{21}$}

\psfrag{Lag - C22}{Lag - $C_{22}$}

\psfrag{Lag - C31}{Lag - $C_{31}$}

\psfrag{Lag - C32}{Lag - $C_{32}$}

\psfrag{Lag - C33}{Lag - $C_{33}$}

\includegraphics[width=6in,height=7in]{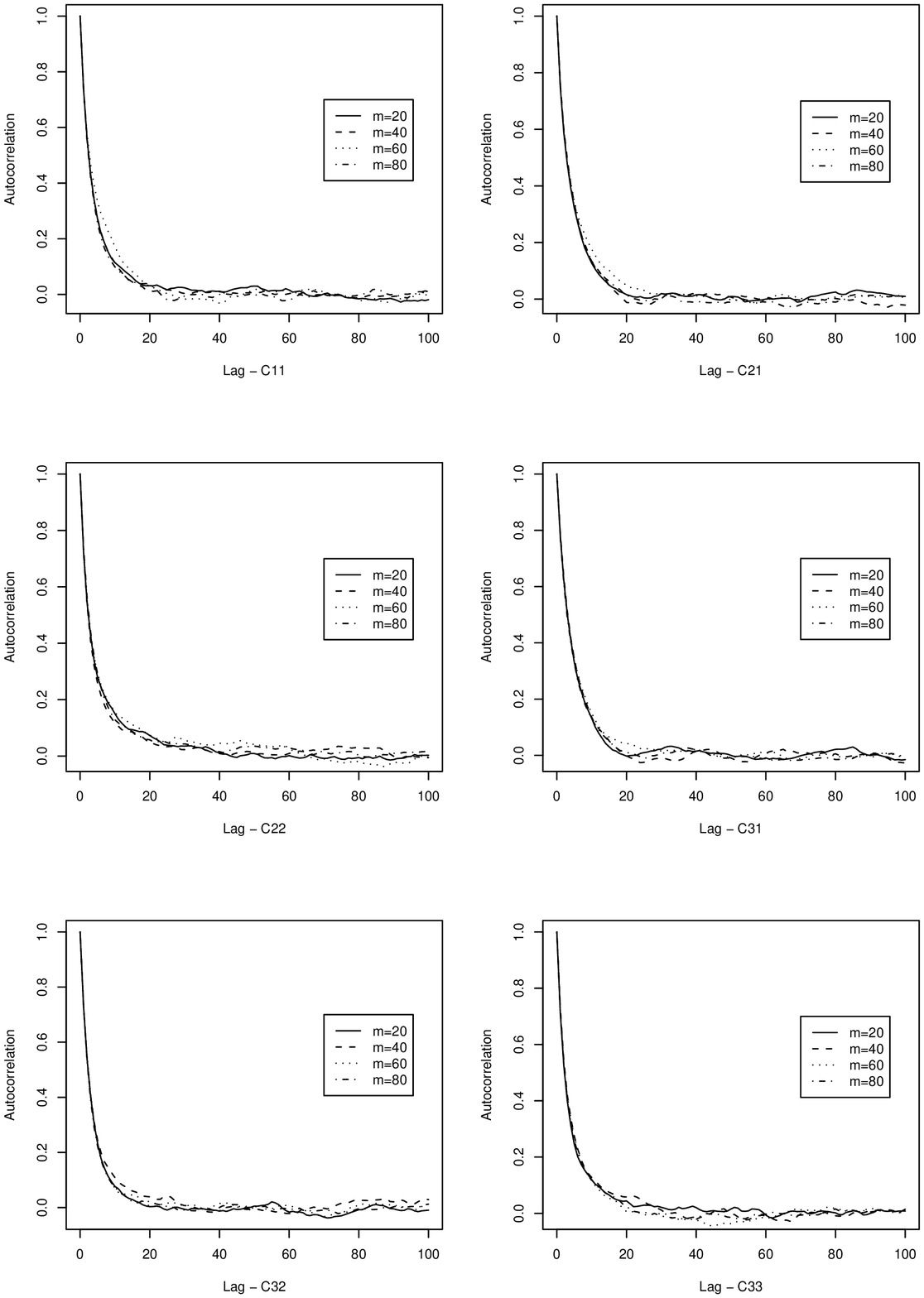}
\caption{Autocorrelation plots for the posterior draws of the $C$
matrix entries for different numbers of imputed points
($m=20,40,60,80$). Simulated data.} \label{simfig1}
\end{figure}

\begin{figure}

\psfrag{mu1}{$\mu_{1}$}

\psfrag{sigma2}{$\sigma_{2}$}

\psfrag{rho3}{$\rho_{32}$}

\includegraphics[width=6in,height=7in]{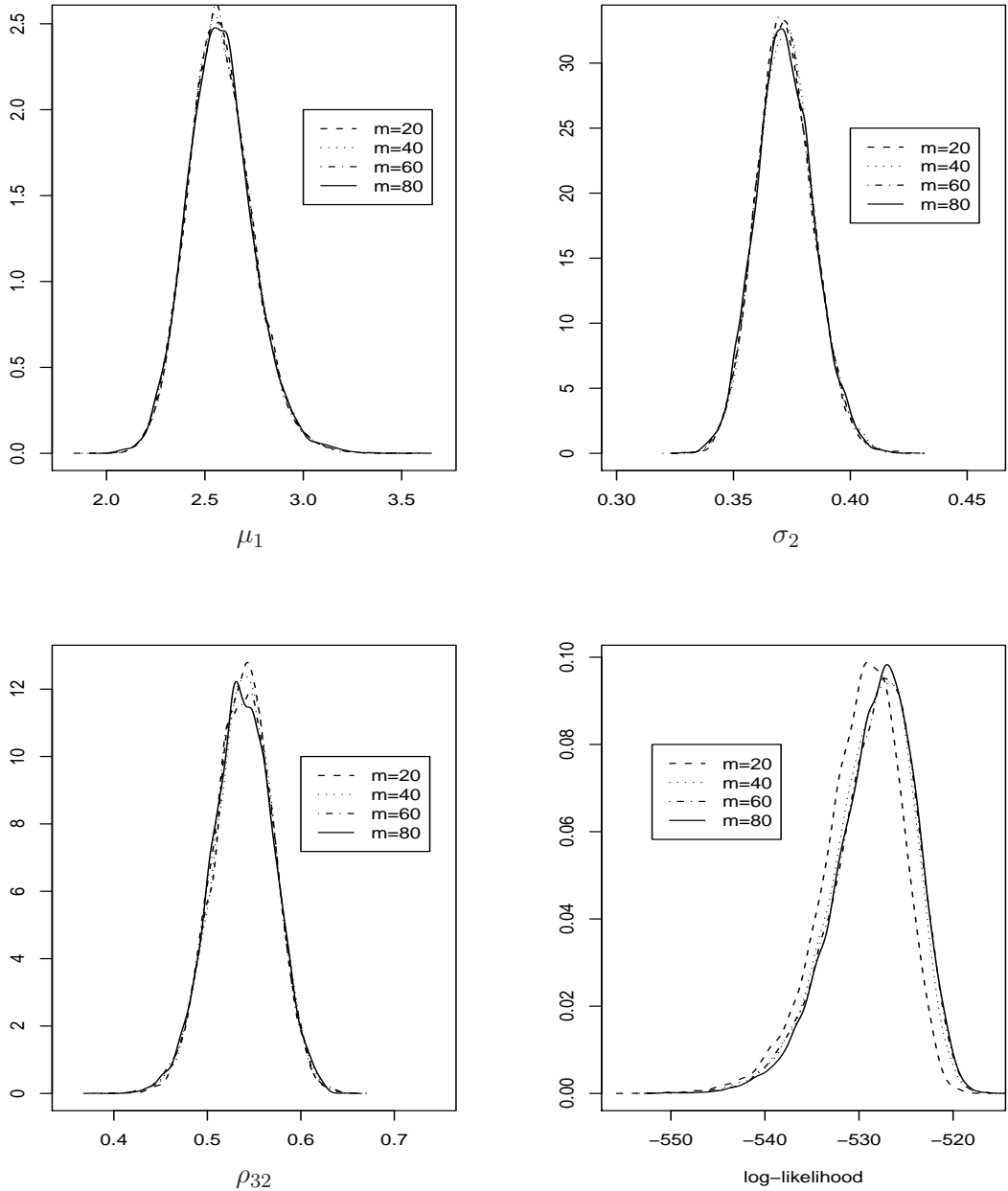}
\caption{Kernel densities of the posterior draws for some
parameters ($\mu_1$, $\sigma_2$, $\rho_{32}$) and the
log-likelihood, for different numbers of imputed points
($m=20,40,60,80$). Simulated data.} \label{simfig2}
\end{figure}

We simulated 500 equidistant observations (apart from the initial
point) at times $\{t_k=k,\;k=0\dots,n\}$ with $t_n=500$. Several
MCMC runs, with different numbers of imputed points
$m$=$\{20,40,60,80\}$, were examined. This was done to monitor the
autocorrelation as well as the approximation error of the
likelihood in relation with the level of augmentation. The
acceptance rate of the independence sampler used for the path
updates was $98.14\%$, raising no concerns regarding its
performance. Figure \ref{simfig1} shows autocorrelation plots for
the posterior draws of the $C$ matrix components. There is no sign
of any increase to raise suspicions against the irreducibility of
the chain. Figure \ref{simfig2} depicts density plots for some
parameters as well as the log-likelihood which may be seen as an
appropriate diagnostic plot for the quality of the approximations.
Densities for $m=60$ and $m=80$ look similar and therefore the
argument that their level of augmentation is sufficient appears to
be plausible. The plots of Figure \ref{simfig2} and the results of
Table \ref{simtab}, which contains summaries of the parameter
posterior draws for $m=80$, are in good agreement with the true
values of the parameters.

\begin{table}
\begin{tabular}{c cccc}
\hline  Parameter  &  True Value & Posterior mean  & Posterior SD
& Posterior median \\\hline
$\kappa_{1}$& 0.2  & 0.174  & 0.025  & 0.174 \\
$\kappa_{2}$& 0.15 & 0.123  & 0.031  & 0.121 \\
$\kappa_{3}$& 0.22 & 0.223  & 0.030  & 0.224 \\
$\mu_{1}$   & 2.5  & 2.578  & 0.167  & 2.571 \\
$\mu_{2}$   & 3.0  & 2.986  & 0.366  & 2.951 \\
$\mu_{3}$   & 2.0  & 1.908  & 0.094  & 1.905 \\
$\sigma_{1}$& 0.45 & 0.434  & 0.016  & 0.434 \\
$\sigma_{2}$& 0.35 & 0.372  & 0.012  & 0.372 \\
$\sigma_{3}$& 0.4  & 0.401  & 0.014  & 0.402 \\
$\rho_{21}$ & 0.45 & 0.480  & 0.034  & 0.480 \\
$\rho_{31}$ & 0.35 & 0.318  & 0.041  & 0.319 \\
$\rho_{32}$ & 0.55 & 0.537  & 0.033  & 0.538 \\ \hline
\end{tabular}

\caption{Summaries of the posterior draws of the model parameters
for $m=80$. Simulated dataset.}\label{simtab}

\end{table}

\section{Application: EUR/USD and GBP/USD exchange rates}
\label{sec:Real Data}

\begin{figure}
\includegraphics[width=6in,height=7in]{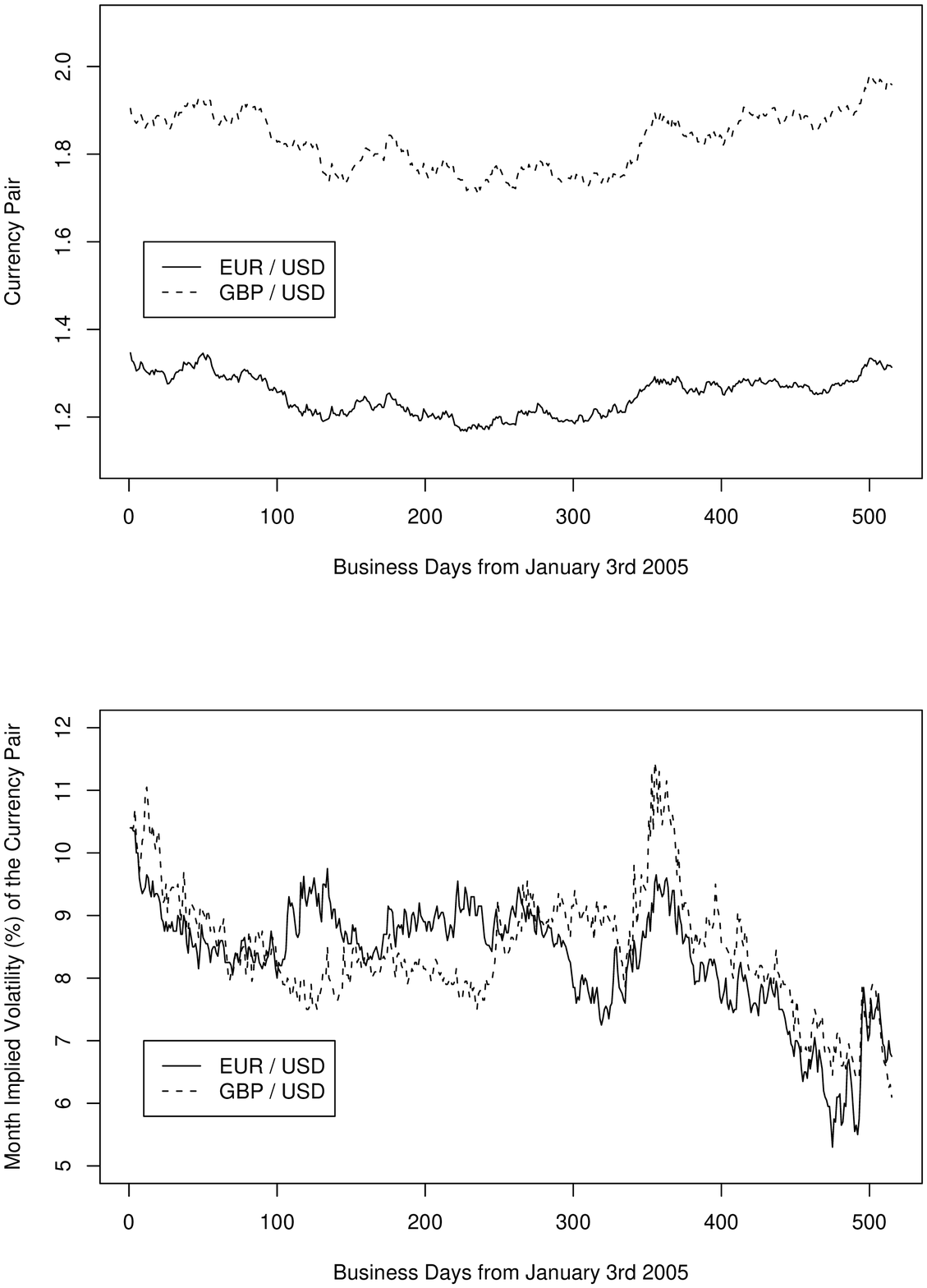}
\caption{Daily EUR/USD and GBP/USD rates (up) and their month
implied volatilities (\%) (down) from 3rd of January 2005 to 22nd
of December 2006.} \label{exfig1}
\end{figure}

The dataset consists of roughly two years of daily exchange
EUR/USD and GBP/USD rates, specifically from the 3rd of January
2005 to 22nd of December 2006. We denote these rates with
$r^{eur/usd}$ and $r^{gbp/usd}$ and their logarithms with
$Y^{eur/usd}$ and $Y^{gbp/usd}$ respectively. Our dataset also
contains the corresponding month implied volatilities constructed
from options made on the currency pairs. The data are plotted in
Figure \ref{exfig1}.

We use the implied volatilities of the currency pairs to construct
proxies for their actual volatilities, denoted with $IV^{eur/usd}$
and $IV^{gbp/usd}$. For simplicity, these proxies are assumed to
be exact observations of the volatilities. Alternative assumptions
are possible, such as their adjustment \citep{ait:kim05}, or a
formulation with noisy observations. Table \ref{extab1} provides
several descriptive statistics including the correlation matrix of
the $4-$dimensional time series containing the implied
volatilities and the log-exchange rates
$Y=\left(IV^{eur/usd},IV^{gbp/usd},Y^{eur/usd},Y^{gbp/usd}\right)$.

\begin{table}
\begin{tabular}{l ccccc}
\hline & Mean & St. Deviation & Median & &\\\hline
$IV^{eur/usd}$$\times$ 100 & 0.693  & 0.076 & 0.708 & &\\
$IV^{gbp/usd}$$\times$ 100 & 0.704  & 0.078 & 0.696 & &\\
$r^{eur/usd}$  & 1.2499 & 0.045   &  1.2578 & &\\
$r^{gbp/usd}$  & 1.8304 & 0.066   &  1.8375 & &\\\hline &
\multicolumn{5}{c}{Correlation Matrix} \\\hline
$\Delta IV^{eur/usd}$ & 1      &        &        & &\\
$\Delta IV^{gbp/usd}$ & 0.5551 & 1      &        & &\\
$\Delta Y^{eur/usd}$  & 0.0148 & 0.0101 & 1      & &\\
$\Delta Y^{gbp/usd}$  & 0.0119 & 0.0075 & 0.8093 & &1\\\hline
\end{tabular}

\caption{Descriptive statistics for EUR/USD and GBP/USD exchange
rates and their implied volatilities.}\label{extab1}

\end{table}

\noindent Note that some correlations appear to be substantial and
should be taken into account in the analysis of the data. Hence we
fit the bivariate Heston model to the $4-$dimensional time series
$Y$ using the MCMC data augmentation scheme of this paper. Section
\ref{ssec:multivariate sv} provides details on the reparametrised
likelihood for the data. For reasons of model parsimony, we only
consider correlations between the pairs
$\left(IV^{eur/usd},IV^{gbp/usd}\right)$ and
$\left(Y^{eur/usd},Y^{gbp/usd}\right)$, and set the remaining ones
($\rho_{31}$,$\rho_{32}$,$\rho_{41}$,$\rho_{42}$) to zero. This is
in line with Table \ref{extab1} and some preliminary analysis
which considered all possible correlations. Note that the
parameters of $C$ that need to be updated are just $C_{11}$,
$C_{21}$, $C_{22}$ and $C_{43}$, as $C_{33}$ and $C_{44}$ are
redundant and the remaining entries are equal to zero like the
corresponding correlations. In other words, there exists a 1-1
mapping between the diffusion matrix elements
($\sigma_1$,$\sigma_2$,$\rho_{21}$,$\rho_{43}$) and
($C_{11}$,$C_{21}$,$C_{22}$,$C_{43}$). We complete the model by
assigning non-informative priors as in the previous section:
$p(\theta)\propto \theta^{-1}$ for the positive parameters
($\kappa_{1}$,$\kappa_{2}$,$\mu_{1}$,$\mu_{2}$,$C_{11}$,$C_{22}$)
and $p(\theta)\propto 1$ for the rest
($\mu_{3}$,$\mu_{4}$,$C_{21}$,$C_{43}$).

\begin{figure}

\psfrag{Lag - C11}{Lag - $C_{11}$}

\psfrag{Lag - C21}{Lag - $C_{21}$}

\psfrag{Lag - C22}{Lag - $C_{22}$}

\psfrag{Lag - C43}{Lag - $C_{43}$}

\includegraphics[width=6in,height=7in]{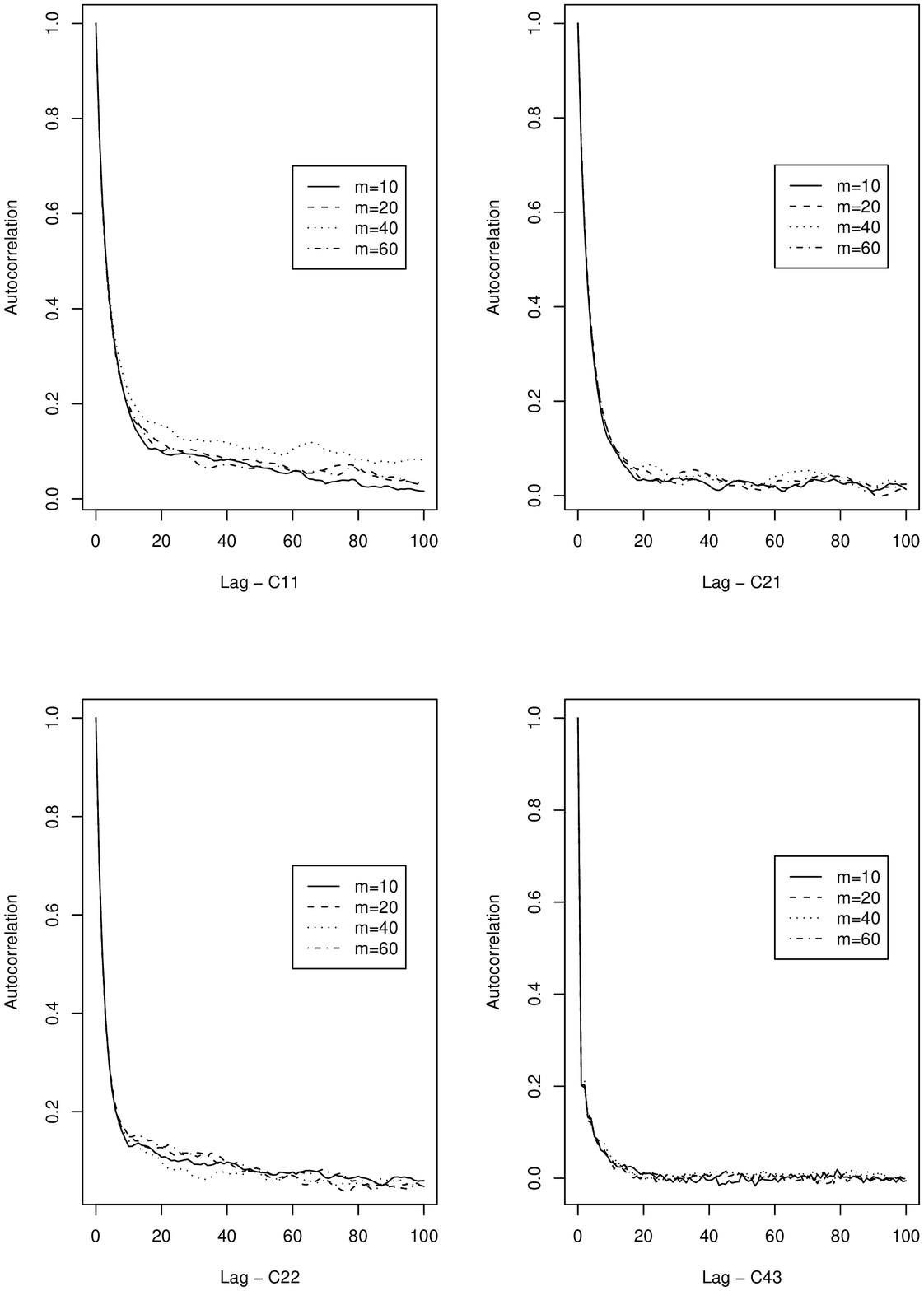}
\caption{Autocorrelation plots for the posterior draws of the $C$
matrix entries for different numbers of imputed points
($m=10,20,40$). EUR/USD and GBP/USD exchange rates dataset.}
\label{exfig2}
\end{figure}

As before, several MCMC runs with different numbers of imputed
points $m$=$\{10,20,40\}$ were used. The data, referring to
business days, were assumed to be equidistant and the time was
measured in years. Again, the acceptance rate of the independence
sampler used for the path updates was particularly high $99.16\%$.
The autocorrelation plots of draws from the posterior of the
parameters $C_{11}$,$C_{21}$,$C_{22}$, and $C_{43}$, in Figure
\ref{exfig2}, reveal no sign of any increase in the level of
augmentation.

\begin{figure}

\psfrag{mu1}{$\mu_{1}$}

\psfrag{mu4}{$\mu_{4}$}

\psfrag{sigma2}{$\sigma_{2}$}

\psfrag{rho1}{$\rho_{21}$}

\psfrag{rho2}{$\rho_{43}$}

\includegraphics[width=6in,height=7in]{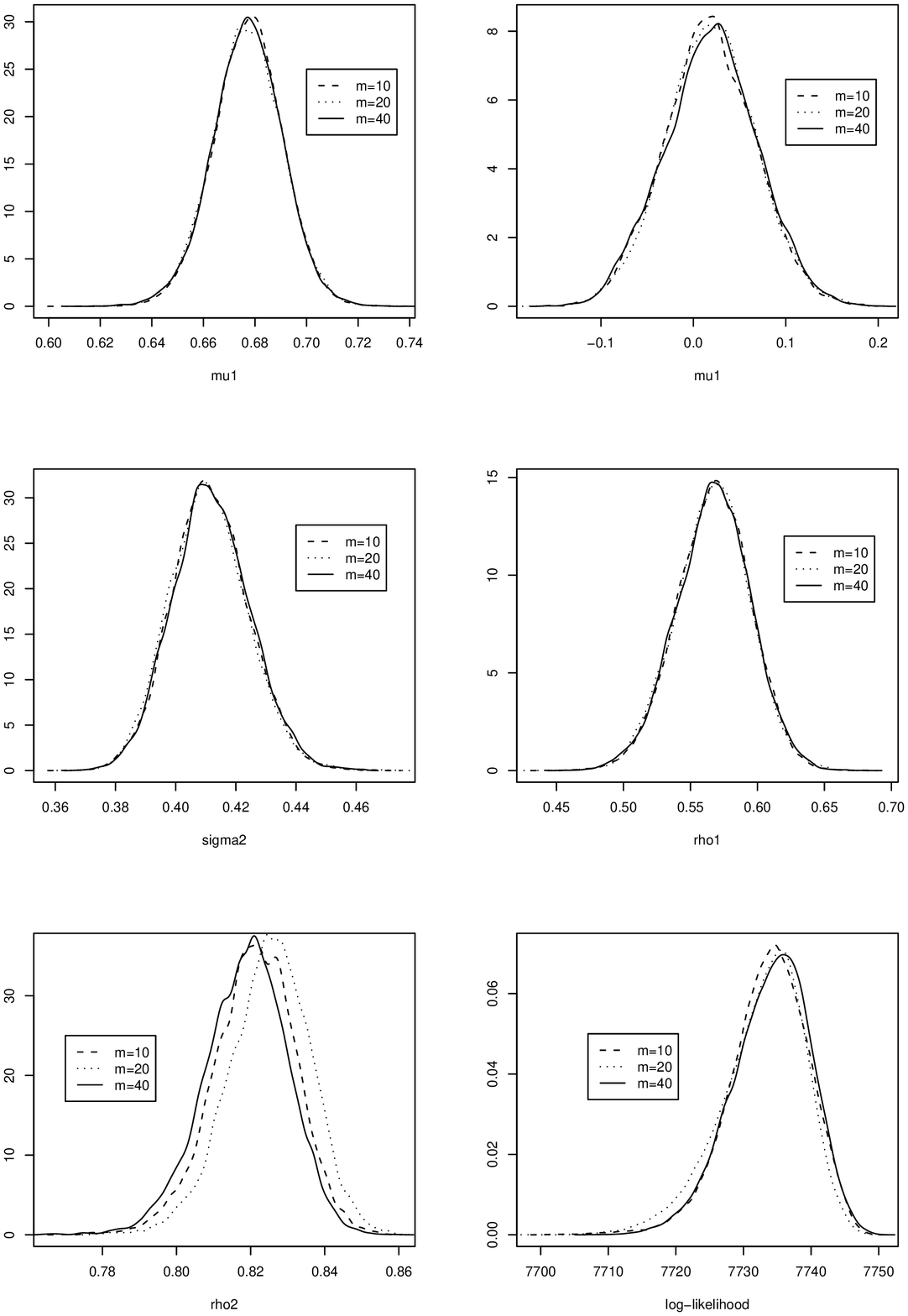}
\caption{Kernel densities of the posterior draws for some
parameters ($\mu_1$, $\mu_1$, $\sigma_2$, $\rho_{21}$,
$\rho_{43}$) and the log-likelihood, for different numbers of
imputed points ($m=10,20,40$). EUR/USD and GBP/USD exchange rates
dataset.} \label{exfig3}
\end{figure}

\noindent Regarding the approximation error due to the
discretisation of the diffusion path, the density plots from the
posterior draws of some parameters and the log-likelihood, in
Figure \ref{exfig3}, provide convergence evidence for the
approximating sequence of the data augmentation scheme.

\noindent Table \ref{extab2} contains summaries of the parameter
posterior draws, where both correlations appear to be high. Note
that the non-parametric estimates of Table \ref{extab1} are based
on the quadratic variation process and are therefore amenable to
bias due to the discretisation of the diffusion path. On the other
hand, the discretisation error of the model estimates may become
arbitrary small. The posterior mean or median values provide point
estimates of the parameters which may be used for option pricing
purposes. Alternatively, the samples from their posterior of the
parameters may be used in a Bayesian option pricing framework. In
any case, it may be useful to take into account the correlated
market structure of the log-exchange rate and their impled
volatilities.

\begin{table}
\begin{tabular}{c  ccc}
\hline  Parameter  &  Posterior mean  & Posterior SD & Posterior
median \\\hline
$\kappa_{1}$ & 0.153  & 0.023  & 0.153  \\
$\kappa_{2}$ & 0.206  & 0.030  & 0.204 \\
$\mu_{1}$ $\times$ 100  & 0.677  & 0.014  & 0.677 \\
$\mu_{2}$ $\times$ 100  & 0.689  & 0.012  & 0.690 \\
$\mu_{3}$    & 0.001  & 0.053  & 0.001 \\
$\mu_{4}$    & 0.019  & 0.049  & 0.019 \\
$\sigma_{1}$$\times$ 100& 0.343  & 0.010  & 0.343 \\
$\sigma_{2}$$\times$ 100& 0.411  & 0.013  & 0.411 \\
$\rho_{21}$  & 0.567  & 0.028  & 0.567 \\
$\rho_{43}$  & 0.821  & 0.011  & 0.821 \\ \hline
\end{tabular}

\caption{Summaries of the posterior draws of the model parameters
for $m=60$. EUR/USD and GBP/USD exchange rates
dataset.}\label{extab2}

\end{table}

\bigskip

\section{Discussion}
\label{sec:discussion}

In this paper we introduced a parametrisation framework based on
the Cholesky decomposition, for handling correlations of
multi-dimensional diffusions in a Bayesian MCMC setting. This
framework facilitates componentwise updates of the diffusion
matrix, in a way so that its positive definite structure is
preserved. It may therefore be of substantial value in high
dimensional diffusion models. The Cholesky factorisation was used
in connection with data augmentation and therefore applies to both
directly and partially observed diffusions. In order to overcome
degenerate MCMC algorithms, the likelihood reparametrisation of
\cite{rob:str01} was generalised to several multi-dimensional
diffusions, including stochastic volatility models, thus providing
a stand alone solution to the problem. Being a data augmentation
scheme, our MCMC algorithm is based on an approximation of the
likelihood, whose error may become arbitrarily small by simply
increasing the level of augmentation.

Nonetheless, the Cholesky factorisation of the diffusion matrix
may be coupled with alternative, to data augmentation, techniques
for approximating the likelihood. The exact inference framework of
\cite{bes:pap:rob:f06} and the analytic likelihood expansions of
\cite{ait05} provide such examples with appealing properties: the
former eliminates entirely the error due to the discretisation of
the diffusion path, whereas the latter provides closed form
expressions of the likelihood. On the other hand, their
generalisation to partially observed diffusion may present major
difficulties.

Apart from the updates of the diffusion matrix parameters, our
MCMC algorithm differs from other data augmentation schemes, such
as those of \cite{ch:pit:she05} and \cite{gol:wil07}, in the
proposal distribution of the independence sampler involved in the
updates of the diffusion paths. Under these schemes, the proposal
may either be the multi-dimensional bridge of the of
\cite{dur:gal02}, or alternatively that of \cite{del:hu07}, with
the target diffusion matrix. Current work investigates the
behavior of all existing approaches in different settings
regarding the dimensionality of the diffusion, the amount of
correlation, and the sparseness of the data.

\section{Acknowledgements}

Part of the work was carried out during a visit to Lancaster
funded through the EU Marie Curie training scheme. The data of
Section \ref{sec:Real Data} were used with the kind permission of
Citigroup.

\bibliography{corrdiff}
\newpage
\appendix

\section{Proofs of propositions }

{\bf Proof of proposition \ref{prop:reducibility}:}

\bigskip

\noindent The proof is based on he reducibility condition of
(\ref{eq:sahalia condition}), for which we need the inverse of
$\Sigma(X_t,\theta)$
$$
\Sigma(X_t,\theta)^{-1}\;=\;(F_{x}(X_t,\theta)\;C)^{-1}\;=\;C^{-1}\;F_{x}(X_t,\theta)^{-1}.
$$

\noindent In coordinate form the above writes

$$
[\Sigma(X_t,\theta)^{-1}]_{ij}=[C^{-1}]_{ij}f^{\{j\}}(x_t^{\{j\}},\theta)^{-1},\;
\forall\; i,j\in\{1,\dots,d\}.
$$

\noindent Hence, it is not hard to see that the reducibility
condition of \cite{ait05} holds because
$$
\frac{\partial [\Sigma(X_t,\theta)^{-1}]_{ij}}{\partial
x_{t}^{\{k\}}}= \frac{\partial
[\Sigma(X_t,\theta)^{-1}]_{ik}}{\partial
x_{t}^{\{j\}}}=0,\;\forall\; i,j,k\in\{1,\dots,d\},\;\text{with
}j<k
$$
\bigskip

\noindent {\bf Proof of proposition \ref{prop:unit vol
transformation}:}

\bigskip

\noindent The diffusion matrix of $U_t$ should be a
$d-$dimensional identity matrix, therefore by Ito's lemma we get

\begin{equation}
\label{eq:proof prop2} \nabla H(X_t,\theta)\; A\; (\nabla
H(X_t,\theta))^{\prime}\;=\;I_d
\end{equation}

\noindent Consider a transformation of the form
$$
H(X_t,\theta)\;=\;B\;G_x(X_t,\theta),
$$

\noindent where $B$ is an arbitrary $d\times d$ matrix,
independent of $X_t$.

\noindent We can write
$$
\nabla H(X_t,\theta)\;=\;B\;D_G(X_t,\theta),
$$

\noindent where $D_G(X_t,\theta)$ is a diagonal matrix with
$$
[D_G(X_t,\theta)]_{ii}=f^{\{i\}}(x_t^{\{i\}},\theta)^{-1},\;i=1,\dots,d.
$$

\noindent Indeed, the $k-$th row of $\nabla H(X_t,\theta)$ equals
$$
\nabla H(X_t,\theta)=\nabla \left(\sum_{j=1}^{d}B_{kj}g^{\{i\}}
(x_t{\{j\},\theta})\right)=
\left(B_{k1},\dots,B_{kd}\right)\;D_G(X_t,\theta).
$$

\noindent If we substitute on (\ref{eq:proof prop2}), using also
(\ref{eq:cholesky matrix}), we get
$$
\;B\;D_G(X_t,\theta)\;F_x(X_t,\theta)\;C\;C^{\prime}\;
F_x(X_t,\theta)\;D_G(X_t,\theta)\;B^{\prime}\;=\;I_d,
$$

\noindent which since $D_G(X_t,\theta)\;F_x(X_t,\theta)\;=
\;F_x(X_t,\theta)\;D_G(X_t,\theta)\;=\;I_d$ becomes
$$
B\;C\;C^{\prime}\;B^{\prime}\;=\;I_d,
$$

\noindent which is satisfied if we set $B=C^{-1}$ .

\end{document}